\documentclass[%
 reprint,
 amsmath,amssymb,
 aps,
pra
]{revtex4-2}

\usepackage{graphicx}

\usepackage{bm}
\usepackage{siunitx}
\usepackage{braket}
\usepackage{tikz}
\usepackage{hyperref}

\begin{document}

\title{Perspective: Quantum gases in bubble traps}

\author{Romain Dubessy}
\altaffiliation[Also at ]{Aix-Marseille University, CNRS UMR 7345, PIIM, 13397, Marseille, France}
\author{Hélène Perrin}
\email{helene.perrin@univ-paris13.fr}
\affiliation{Université Sorbonne Paris Nord, Laboratoire de Physique des Lasers, CNRS UMR 7538, 99 av. J.-B. Clément, F‐93430 Villetaneuse, France}

\begin{abstract}
This paper presents a review and perspective on quantum gases in bubble traps. We emphasize how the idea of realizing shell-shaped condensates emerged and was enabled by the invention of the radiofrequency adiabatic potential technique. We review the many subsequent theoretical works that address the new physics emerging for a condensate trapped on a closed surface. We present the current status of the experiments, the challenges ahead and highlight how a different approach using an immiscible mixture of two condensates enabled the first observation of a shell-shaped degenerate gas. Finally we list a few open questions that we believe provide interesting research directions.
\end{abstract}

\maketitle

\section{Context, theoretical interest and first experiments with bubbles}
\label{sec:context}

The first observation of a Bose-Einstein condensate (BEC) in an atomic gas featured rubidium atoms confined in a harmonic trap \cite{Anderson1995}. Since then, quantum gases have been prepared in a variety of confinement geometries realized with magnetic or optical trapping, including optical lattices \cite{Bloch2005}, low dimensional harmonic traps \cite{Goerlitz2001,Richard2003,Hellweg2003,QGLD2003,Laburthe2004a, Kinoshita2004,Paredes2004a,Hammes2003,Hadzibabic2006}, double wells \cite{Albiez2005,Schumm2005b}, box potentials \cite{Gaunt2013}, ring traps \cite{Gupta2005,Heathcote2008,Ramanathan2011,Sherlock2011,Navez2016,Ryu2007,Ryu2010,Moulder2012,Eckel2014b,Corman2014} or more sophisticated potentials in the context of atomtronics \cite{Henderson2009,Ryu2015,Brantut2012,Gauthier2016,Gauthier2021,Amico2021,Amico2022}. A new trapping geometry gives access to new physical phenomena, including enhanced correlations \cite{Bloch2008RMP}, Josephson oscillations \cite{Albiez2005}, or persistent currents enabled by the annular topology \cite{Ryu2007,Ryu2010,Moulder2012,Eckel2014b,Corman2014}.

Another interesting topology is the one of the sphere \cite{Lundblad2023,Tononi2024a}. When a quantum gas is strongly confined to a closed surface such as a sphere or an ellipsoid, the system is compact in the mathematical sense, with specific boundary conditions. On a sphere, all points are alike in the absence of symmetry breaking by axial rotation or a magnetic field, which realizes a two-dimensional rotationally invariant system with homogeneous curvature. In an ellipsoid, the curvature is non uniform and can lead to local confinement of atoms near the higher curvature regions \cite{DaCosta1981,Kaplan1997,Sandin2017a}. The curvature also influences collisional properties \cite{Zhang2018,Tononi2022,Shi2017} and can stabilize solitons for particles with attractive interactions \cite{Tononi2024c}.

Beyond a modification of the critical temperature for Bose-Einstein condensation or superfluidity \cite{Bereta2019,Tononi2019}, superfluid dynamics on a sphere presents unique properties with respect to the case of flat gases. Specific collective modes characterize a spherical superfluid \cite{Lannert2007,Sun2018}. On a closed surface, due to neutrality of the total vortex charge, vortices are always present by pairs \cite{Turner2010}. The compactness of the closed surface can also lead to long-range interactions between vortices, depending on its aspect ratio \cite{Turner2010}. A closed surface also gives a natural analogy to geophysical flows \cite{Skipp2022} or astrophysical systems \cite{Verma2022}. While the first theoretical works focused on superfluid helium, dynamics of vortices in superfluids confined on closed surfaces has recently been the subject of several theoretical studies in the context of ultracold atoms \cite{Bereta2021,Caracanhas2022,Tononi2024b}. A recent paper gives an accurate review of the current knowledge on this topic to date \cite{Tononi2024a}. Finally, strong excitation of the quantum gas can lead to a turbulent state, where either vortices or wave-like excitations can be present. Turbulence has been studied with quantum gases in harmonic traps \cite{Henn2009b}, and more recently in boxes with a uniform initial density \cite{Navon2016,Galka2022}. In the case of turbulence on a curved surface, we can expect direct analogies with a turbulent atmosphere \cite{Skipp2022,Muller2024}. 
The possibility to get information on phase patterns in an interferometric measurement in a double bubbles is promising for investigating this topic, as discussed in Ref. \cite{Beregi2024} of this Special Issue.

On the experimental side, an efficient method to confine ultracold atoms on a closed surface was proposed in 2000 by Oliver Zobay and Barry Garraway \cite{Zobay2000,Zobay2001,Zobay2004}. The basic idea relies on adiabatic potentials arising from the resonant interaction of atoms placed in an inhomogeneous magnetic field with a radiofrequency (rf) field. At the positions in space where the rf field is resonant with the splitting between the magnetic sublevels due to the Zeeman interaction, the local eigenstates dressed by the rf field are split by the rf coupling. If the atoms follow adiabatically the local eigenstate as they move across this resonance, it results in adiabatic potentials which present either a local maximum or a local minimum at the resonance points, depending on the dressed magnetic sublevel. For the upper level, this configuration provides a trapping potential that is minimum on the isomagnetic surface resonant with the rf field. Around a minimum of the static magnetic field, this resonant surface is a `bubble' or `shell', with the topology of a sphere. For a comprehensive review of the adiabatic potential technique using rf-dressing, see Ref. \cite{Garraway2016,Perrin2017}.

\begin{figure}[t]
    \centering
    \includegraphics[width=0.5\linewidth]{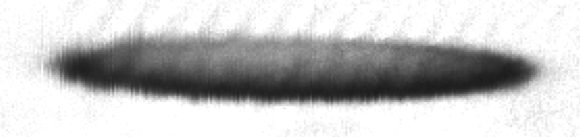}
    \caption{Atomic density distribution of an ultracold gas confined at the surface of an ellipsoidal bubble trap of radii $\SI{160}{\micro\metre}\times\SI{1.3}{\milli\metre}$, obtained by dressing atoms in a Ioffe-Pritchard trap. Reused with permission from O. Morizot, `Pi\`eges radiofr\'equence tr\`es anisotropes pour un condensat de Bose-Einstein', Ph.D. thesis, Universit\'e Paris 13 (2007) \cite{MorizotThese}. First published
in Ref. \cite{Colombe2004b} with a different color map.
    \label{fig:bubble_Morizot}
    }
\end{figure}

The first experimental demonstration of bubble traps relying on this method dates back from 2003 \cite{Colombe2004b}. A cigar-shaped Bose-Einstein condensate was loaded from a Quadrupole and Ioffe Configuration (QUIC) magnetic trap \cite{Esslinger1998} into an ellipsoidal bubble trap with a marked anisotropy \cite{Colombe2004a,MorizotThese}, see Fig.~\ref{fig:bubble_Morizot}, resulting in a cloud of ultracold atoms at the bottom of the ellipsoid. The transfer of the dressed atoms to a dressed quadrupole trap led to a bubble with a smaller anisotropy \cite{Morizot2007}.

In the early realizations, the ultracold atomic cloud in the bubble potential was in a thermal state, even if the gas was condensed before transfer to the dressed state \cite{Colombe2004a,White2006}. It was rapidly understood that the quality of the rf source was essential to avoid heating and losses during and after transfer, in particular its amplitude and frequency noise characteristics \cite{Morizot2008}. Important experimental developments on magnetic field stability and the application of Direct Digital Synthesis (DDS) rf sources to the manipulation of ultracold atoms \cite{Fletcher2008} also benefited to rf adiabatic potentials \cite{Gildemeister2010} and allowed the production of degenerate Bose gases at the bottom of rf-dressed quadrupole bubble traps \cite{Gildemeister2012,Merloti2013a}. The current experiments benefit from 20 years of experimental developments and enable a high level of control of the static and dynamic geometry of rf-dressed adiabatic potentials.

Radiofrequency dressing for adiabatic potentials \cite{Garraway2016,Perrin2017} is a powerful technique and also give access to other topologically interesting situations. Double wells were obtained with atom chip devices \cite{Schumm2005b,Extavour2006,Jo2007a}, enabling the study of phase fluctuations in elongated quantum gases \cite{Schumm2005b,Hofferberth2007b,Jo2007c}, and more recently in quasi-two-dimensional Bose gases \cite{Barker2020,Sunami2022,Sunami2023,Beregi2024} using atom dressing with multiple rf frequencies \cite{Harte2018,Mas2019}. Ring traps can be achieved by intersecting an ellipsoidal bubble near its equator with an optical dipole potential \cite{Morizot2006,Heathcote2008,deGoer2021}, by a specific choice of rf field polarization and static field configuration \cite{Lesanovsky2006a,Kim2016,Guo2022} or by a modulation of the magnetic and rf fields to create so-called time-averaged adiabatic potentials (TAAP) \cite{Lesanovsky2007b,Sherlock2011,Navez2016}. More generally, modulation of the trap parameters opens the way to a larger variety or trapping geometries \cite{Lesanovsky2007b,Gildemeister2010} and was used to induce superfluid dynamics at the bottom of bubble traps, for the study of collective modes of quasi two-dimensional Bose gases \cite{Dubessy2014,DeRossi2016,Merloti2013b}, rotation in ring traps \cite{Navez2016,Pandey2019}, or fast rotation directly in the bubble, yielding vortex lattices \cite{Gildemeister2012,Sharma2024} or a dynamical ring sustained by its angular momentum \cite{Guo2020}.

These last results were obtained in a partially filled bubble trap, either at the equator in the case of rings or at the bottom for quasi-two-dimensional Bose gases. Filling the bubble requires a gas with an overall energy larger than the potential inhomogeneity along the surface, which on Earth includes gravity. An early example is presented in Fig.~\ref{fig:bubble_Morizot}, with a thermal gas of thermal energy larger than the difference in gravitational energy between the top and the bottom of the bubble based on a dressed Ioffe-Pritchard magnetic trap \cite{MorizotThese}. Reaching the same situation with a degenerate quantum gas confined in adiabatic potentials is a challenge that would open a wide range of possible experiments. It requires in particular to compensate for gravity, which motivated an experiment performed at the International Space Station (ISS) \cite{Carollo2022}. In the next section, we discuss the various proposals and the advancement of current experiments toward this goal.

\section{The quest for filled bubbles}

The recent attempts to load a bubble-shaped trap in the Cold Atom Laboratory aboard the International Space Station has attracted much interest and triggered several theoretical investigations aiming at an accurate description of the properties of a shell-shaped Bose-Einstein condensate. While the initial proposal focused more on discussing the geometrical properties and the feasibility of this novel kind of traps~\cite{Zobay2000,Zobay2001,Zobay2004}, it was soon realized that this trap geometry would also modify the equilibrium and the dynamical properties of the condensate~\cite{Lannert2007} with respect to the standard harmonic trap geometry. In particular, the fact that the condensate is hollow gives rise to new types of collective modes and a non trivial time-of-flight expansion, exhibiting strong self interference~\cite{Lannert2007}.

\subsection{Theoretical tools to describe bubble-shaped superfluids}
\label{sec:theory}

A number of papers addressed the question of the critical temperature to achieve condensation in a bubble trap~\cite{Bereta2019,Tononi2019,Tononi2020,Rhyno2021}. At the level of the ideal gas, the spherical geometry changes the density of states and tends to lower the critical temperature~\cite{Bereta2019}. The density of states can be evaluated analytically, taking into account finite size effects, or using a semi-classical approach, which gives accurate results for large enough radii~\cite{Bereta2019}. When considering the strict two-dimensional limit, in which atoms live on the surface of a sphere, it is still possible to define a critical temperature for Bose-Einstein condensation, owing to the finite size of the system. However, when contact interaction between the atoms is added, one recovers a Kosterlitz-Thouless transition to a superfluid state~\cite{Tononi2019} for a finite size system. When a more realistic three-dimensional shell potential is taken into account, it is usually sufficient to estimate the ideal gas BEC critical temperature using a semiclassical approach~\cite{Bereta2019,Tononi2020} and, if needed, Quantum Monte-Carlo algorithms can be adapted to study mesoscopic atom ensembles in this geometry~\cite{Tononi2020}.
Finally, it is worth noting that in realistic experiments the condensate is usually first obtained in a standard harmonic trap, before being transferred to the shell-shaped trap. If this transfer is done adiabatically, the temperature is expected to decrease, due to the expansion, however the critical temperature drops even faster and the scaling is not favorable to preserve the condensate~\cite{Rhyno2021}. This suggests that a fine tuning of the trap as well as active cooling, for instance using rf-evaporation, might be necessary to achieve a hollow bubble condensate. The Paris group recently managed to load a hemisphere of a shell potential with a BEC using a controlled expansion while compensating gravity in an Earth-based laboratory~\cite{Guo2022}.

\begin{figure}
    \centering
    \includegraphics[width=8.6cm]{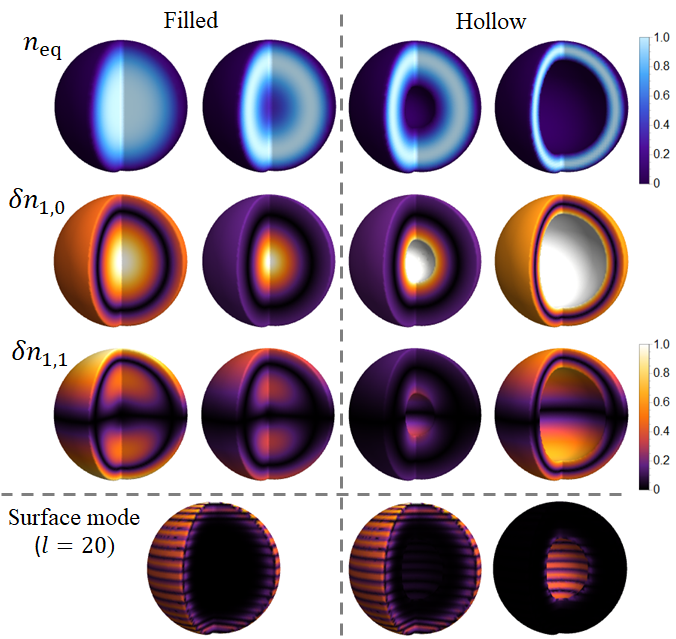}
    \caption{Illustration of the hollowing-out transition from a filled sphere to a spherical surface.
    First row: equilibrium density profile, following rows: density perturbations for low energy Bogoliubov modes. The last row evidences the new branch of surface modes localized at the inner surface, which is unique to hollow geometries.
    Reused from K. Sun, K. Padavi\'c, F. Yang, S. Vishveshwara, and C. Lannert, Phys. Rev. A 98, 013609 (2018) \cite{Sun2018}. Copyright (2018) by the American Physical Society.
    \label{fig:hollowing-out}
}
\end{figure}

Apart from the evolution of the critical temperature, it is interesting to look for other signatures of the hollowing-out transition when a quantum gas is expanded from a simple harmonic trap to a shell. As mentioned before, the time-of-flight expansion of a spherical cloud acquires new features and displays a sharp self-interference peak at zero momentum~\cite{Lannert2007,Tononi2020}. Another striking feature can be observed in the collective mode spectrum of the condensate: due to the inner surface, new branches emerge for a hollow condensate~\cite{Lannert2007,Padavic2017,Sun2018}, as illustrated in Fig.~\ref{fig:hollowing-out} for the spherical symmetry, and the breathing mode displays a minimum value at the hollowing-out transition~\cite{Padavic2017,Sun2018}.
In this context it is natural to raise the question of the effective dimensionality of the system. Indeed starting from a three-dimensional simply connected BEC, the expansion in a shell geometry will occur through a thick shell configuration before entering a quasi-two-dimensional regime, for large enough radii, where the radial motion will be essentially frozen. This impacts the critical temperature~\cite{Bereta2019} as well as the nature of the superfluid transition~\cite{Tononi2019} and the collective mode spectrum~\cite{Sun2018}. The dimensionality reduction on a curved manifold must be carried out carefully~\cite{Moller2020,Biral2024}, especially when the curvature or the confinement to the surface are not uniform~\cite{Guo2022}.

\begin{figure}
    \centering
    \includegraphics[width=6cm]{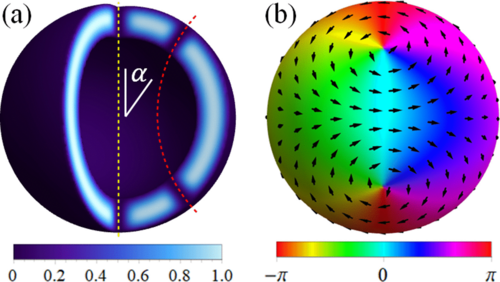}
    \caption{(a) Density profile of a shell-shaped condensate with a pair of vortices. (b) Phase of the condensate wavefunction on the shell, encoded into the colormap. The small arrows give the direction of the local velocities showing the superfluid flow. 
    Reused from K. Padavi\'c, K. Sun, C. Lannert, and S. Vishveshwara, Phys. Rev. A 102, 043305 (2020) \cite{Padavic2020}. Copyright (2020) by the American Physical Society. 
    \label{fig:2vortices}
    }
\end{figure}

The sphere or shell geometry is particularly interesting to study quantum vortex physics, due to its closed surface topology: in fact vortices must come by pairs, as illustrated in Fig.~\ref{fig:2vortices}, which leads to new configurations and dynamics~\cite{Turner2010,Padavic2020,Bereta2021,Rhyno2021,Caracanhas2022,Li2023,Saito2023,White2024,Xiong2024}, with respect to open surface geometries.
The study of vortex dynamics is greatly simplified by dimensionality reduction: when the superfluid is restricted to the surface of a sphere, one can use a stereographic projection on a plane~\cite{Bereta2021} to obtain an analytical description of the dynamics, using point-like vortex models. This procedure can be extended in principle to any closed surface geometry using a conformal map to the complex plane~\cite{Caracanhas2022}. When many interacting vortices are present, it is possible to use a coarse grained approximation to study the large scale flows, resulting in a vortex fluid model on a closed surface~\cite{Xiong2024}.
A more realistic picture can be obtained by a full numerical simulation of the Gross-Pitaevskii equation for a shell-shaped gas in a rotating frame, evidencing the competition between curvature, rotation and dimensionality~\cite{Padavic2020,White2024}. For fast rotations in a rotationally invariant bubble, a transition to a giant vortex state is predicted~\cite{White2024}, reminiscent of what has been observed for harmonic plus quartic potentials~\cite{Fetter2001,Kasamatsu2002,Guo2020}.
Interestingly, the large scale flows emerging when many vortex dipoles are present on a closed surface are similar to geophysical flows~\cite{Skipp2022}, including Kelvin, Yanai, Poincaré equatorial modes~\cite{Li2023} and possibly Rossby-Haurwitz waves~\cite{Saito2023,Xiong2024}.

Finally, the physics of shell-shaped condensates can acquire even richer features by allowing for a spin-degree of freedom \cite{Andriati2021}. For example, a three components spinor BEC on a spherical shell is expected to display new types of solitons, called lump-soliton, described by a winding number associated with a topological feature shared between real and spin spaces~\cite{He2023b}. In the case of a two component spinor BEC, with an additional field coupling the two components (for example an rf signal) and modulating the inter-component interaction, the miscible phase supports Faraday waves that compete with the phase separation mechanism~\cite{Brito2023}, leading eventually to a time-periodic behavior due to the sphere topology. In the case of a BEC with long-range dipolar interactions in a closed geometry, the anisotropic nature of the dipolar interaction will lead to a groundstate exhibiting a density modulation~\cite{Adhikari2012}, which can be used to counteract other inhomogeneities, such as a small residual gravitational potential~\cite{Arazo2021}. In the limit where the dipolar interaction dominates over the contact interaction, a transition to a supersolid phase is expected \cite{Ciardi2024,Ghosh2024}. The experimental implementation of these theoretical proposals with dipolar gases would ideally require a nonmagnetic trap to enable an independent control of the magnetic field, which is not the case with rf-dressed traps.

\subsection{Challenges for rf-dressed trap experiments}
\label{sec:challenges}
As was already underlined in the seminal work~\cite{Zobay2004}, a BEC can spread out on the whole surface of a bubble-shaped trap only if the detrimental contribution of the gravitational potential is mitigated by another effect. We may distinguish two different strategies: overcome the gravitational potential or reduce its influence. The former strategy is easy to implement for a thermal gas, as was done in the first realization of a bubble-shaped trap~\cite{MorizotThese}, whereas the later requires either to reduce the real gravity field by using for example an Earth-orbiting laboratory~\cite{Aveline2020}, or an external field compensating gravity~\cite{Zobay2004,Guo2022}.

To discuss the relevant orders of magnitude, let us consider a spherical trap model for the condensate, with isotropic harmonic confinement to the surface, described by the stationary Gross-Pitaevskii equation:
\begin{eqnarray}
\mu\psi(\bm{r})&=&\left[-\frac{\hbar^2\bm{\nabla}^2}{2M}+U|\psi(\bm{r})|^2\right.\\
\nonumber&&\left.+\frac{M\omega_r^2}{2}(r-r_0)^2+Mgr\left(1+\cos{\theta}\right)\right]\psi(\bm{r}),
\label{eqn:GPE}
\end{eqnarray}
where $\mu$ is the chemical potential, $M$ the atom's mass, $\omega_r$ the harmonic confinement, $r_0$ the radius of the spherical bubble surface, $g$ the gravitational acceleration, $U=4\pi \hbar^2 a_s/M$ the usual interaction strength with $a_s$ the $s$-wave scattering length and $(r,\theta,\phi)$ the spherical coordinates. This model is relevant if the radius of the bubble is much larger than the typical confinement scale: $r_0\gg a_r=\sqrt{\hbar/(M\omega_r)}$.

According to Eq.~\eqref{eqn:GPE}, the condensate will fill the surface of the sphere if the chemical potential overcomes the gravitational potential energy difference between the bottom and the top of the sphere, satisfying $\mu>2Mgr_0$. If we require that the condensate is in the quasi-two dimensional regime, we have also $\mu<\hbar\omega_r$, which imposes a minimal value for the transverse confinement frequency: $\omega_{\rm min}=2Mgr_0/\hbar$, close to $\omega_{\rm min}=2\pi\times\SI{43}{kHz}$ for a sphere with $r_0=\SI{10}{\micro\metre}$ filled with a $^{87}$Rb quantum gas. This is more than an order of magnitude larger that any current experimental realization of shell potentials. It is therefore necessary to compensate gravity in order to realize a quasi-two dimensional superfluid with the topology of a sphere.

In the opposite limit where the cloud is in the three-dimensional regime everywhere on the sphere, we may solve Eq.~\eqref{eqn:GPE} using a three-dimensional Thomas-Fermi solution as follows:
\begin{equation}
n(\bm{r})=|\psi(\bm{r})|^2=\frac{\mu-Mgr_0(1+\cos{\theta})-\frac{M\omega_r^2}{2}(r-r_0)^2}{U},
\label{eqn:dens}
\end{equation}
where we have neglected the gravitational sag, of order $g/\omega_r^2$.
By integrating Eq.~\eqref{eqn:dens} and using the minimal value of the chemical potential $\mu_{\rm min}=2Mgr_0$, we obtain an explicit expression for the minimum atom number required to fill the surface:
\[
N>N_{\rm min}=\frac{32}{105}\frac{g^{3/2}M^2r_0^{5/2}(4g+7r_0\omega_r^2)}{a_s\hbar^2\omega_r^3}.
\]
For a $^{87}$Rb quantum gas, $r_0=\SI{10}{\micro\metre}$ and a moderate transverse confinement frequency $\omega_r=2\pi\times\SI{1}{kHz}$ we find $N_{\rm min}=1.2\times10^7$, which is again very challenging to reach with the adiabatic potential technique. Therefore, it is highly relevant to reduce the effect of gravity \cite{Carollo2022,Guo2022} to produce a shell-shaped quantum gas.

In the case of a thermal gas, it is sufficient to have a thermal energy exceeding the gravitational energy difference between the south and north poles of the bubble, $k_BT>2Mgr_0$, leading to a few microkelvins for a $^{87}$Rb gas in a bubble with radius $r_0=\SI{10}{\micro\metre}$. Gravity compensation or microgravity is nevertheless also necessary for the study of curvature effects on collisions \cite{Zhang2018,Tononi2022,Shi2017}, as these effects appear at very low temperature.

\begin{figure}
    \centering
    \includegraphics[width=6cm]{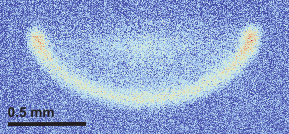} 
    \caption{Ultracold thermal atom gas loaded in an rf dressed adiabatic potential in a micro-gravity environment. Due to residual rf coupling inhomogeneity, only one hemisphere of the shell is loaded. Note the large scale of the shell structure, indicated by the reference scale, enabled by microgravity. 
    Adapted with permission from R. A. Carollo, D. C. Aveline, B. Rhyno, S. Vishveshwara, C. Lannert, J. D. Murphree, E. R. Elliott, J. R. Williams, R. J. Thompson, and N. Lundblad, Nature 606, 281-286 (2022) \cite{Carollo2022}. Copyright (2022), Springer Nature.
    \label{fig:ISS_bubble}
}
\end{figure}

As mentioned at the end of Sec.~\ref{sec:context}, it has been already demonstrated that it is possible to load an rf-dressed Ioffe-Pritchard trap with thermal atoms that cover the entire surface, see Fig.~\ref{fig:bubble_Morizot}. However, to study lower temperature systems and eventually BECs, it is necessary to reduce the magnitude of gravity. This is possible for example in the BEC-CAL facility~\cite{Aveline2020,Frye2021}, in which an ultracold atom machine aboard the ISS enables the study of BECs in micro-gravity environment~\cite{Gaaloul2022,Carollo2022,Oudrhiri2023}. However it remains an experimental challenge to fill a bubble-shaped trap with a degenerate gas even in microgravity~\cite{Lundblad2019}, see Fig.~\ref{fig:ISS_bubble}. Indeed one has to control extremely well the inhomogeneities of the potential, such that the energy variations on the surface are smaller than the chemical potential. For a sphere of radius \SI{10}{\micro\metre}, with a transverse oscillation frequency $\omega_r=2\pi\times\SI{1}{\kilo\hertz}$ filled with $10^5$ $^{87}$Rb atoms, the gas is in the two-dimensional regime and we have $\mu-\hbar\omega_r/2\simeq h\times\SI{720}{Hz}$, which sets the maximum energy scale for potential inhomogeneity.

Among the possible sources of inhomogeneity, apart from gravity, the main one for rf-dressed traps is due to the inhomogeneity of the rf coupling $\Omega_0$, induced by rf field gradients or change in the quantization axis orientation across the surface~\cite{Lundblad2019,Guo2022}. This effect is quite important because adiabatic potentials typically use rf coupling amplitudes of the order of tens of kilohertz: then a variation of only one percent of the coupling on the surface results in a modulation of the trap bottom by hundreds of hertz, thus at the level of $\tilde{\mu}$. Assuming homogeneous rf coupling, the next leading order inhomogeneity comes from variations of $\omega_r$ that also induce a direct modulation of the trap zero-point energy~\cite{Guo2022}.
This effect is smaller because for preserving adiabaticity rf-dressed traps operate in the regime $\omega_r\ll\Omega_0$ \cite{Perrin2017}.
Transverse frequency inhomogeneity may occur in particular if the magnetic field gradient $b^\prime$ is non uniform as $\omega_r\propto b^\prime/\sqrt{\Omega_0}$. Finally, if both $\Omega_0$ and $\omega_r$ are uniform on the surface, a remaining inhomogeneity will be caused by the variation of the local radius of curvature $R$, inducing potential modulations of the order of $\hbar^2/MR^2\approx\hbar^2/Mr_0^2$, which corresponds to a very small energy scale of the order of a few hertz~\cite{Guo2022}.

\begin{figure}
    \centering
    \includegraphics[width=6cm]{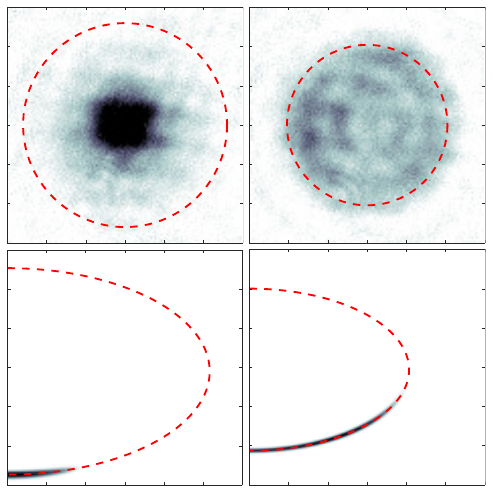} 
    \caption{Quantum gas loaded in an rf dressed shell potential in an Earth-based laboratory. The top row shows in-situ absorption images of the quantum gas (top-view), the dashed red circle being the shell equator, and the bottom row the result of a numerical simulation of the Gross-Pitaevskii equation (side view), the dashed red curve indicating the shell surface. The left column shows the gas at equilibrium at the bottom of the shell when gravity dominates. The right column shows the new groundstate after partial gravity compensation, where the quantum gas fills one hemisphere of the bubble. 
    Reused with permisson from Y. Guo, E. Mercado Gutierrez, D. Rey, T. Badr, A. P. Perrin, L. Longchambon, V. S. Bagnato, H. Perrin, and R. Dubessy, New Journal of Physics 24, 093040 (2022) \cite{Guo2022}. Copyright (2022) under a Creative Commons License.
    \label{fig:LPL}
}
\end{figure}

Radiofrequency inhomogeneity can also turn out to be useful. It was recently demonstrated that the strong rf-coupling inhomogeneity in an rf-dressed quadrupole trap could be turned into an efficient mean of compensating the effect of gravity on the shell surface~\cite{Guo2022}, enabling the observation of a condensate filling the southern hemisphere of a bubble trap, as shown in Fig.~\ref{fig:LPL}. Even if this cannot lead to the realization of a full bubble~\cite{Guo2022}, it allows to study in an Earth-based laboratory the next-order effects limiting the homogeneity of the bubble trap: rf amplitude gradients, rf polarization control, non-uniform transverse confinement, and possibly curvature induced effects. We envision that it will help designing mitigating techniques to control these effects, which could be adapted to micro-gravity setups.

\subsection{The first bubble condensate}
\begin{figure}
    \centering
    \includegraphics[width=6cm]{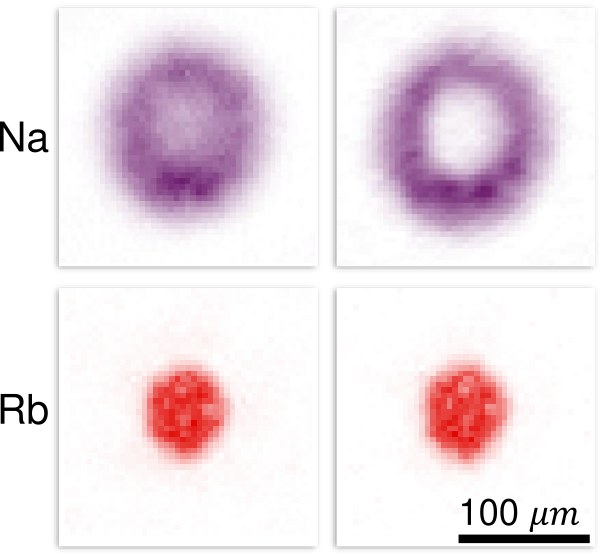}
    \caption{Time-of-flight absorption imaging of a dual species quantum gas mixture, where the lighter element (Na) forms a thin shell around the heavier one (Rb). The right column highlights the hollow shape of the Na condensate, by imaging only the equatorial plane of the bubble. Note that the sizes are magnified by the time-of-flight expansion, the in-situ sizes being approximately five times smaller.
    Adapted from F. Jia, Z. Huang, L. Qiu, R. Zhou, Y. Yan, and D. Wang, Phys. Rev. Lett. 129, 243402 (2022) \cite{Jia2022}. Copyright (2022) by the American Physical Society. 
    \label{fig:mix_bubble}
}
\end{figure}

The first experimental observation of a bubble-shaped quantum gas with the topology of a sphere was recently achieved~\cite{Jia2022}, using a completely different technique, see Fig.~\ref{fig:mix_bubble}. It relies on the idea that a mixture of two mass imbalanced quantum gases held in the same optical trap can phase separate radially, leading to a core filled with the heaviest species while the lighter one forms a thin shell around~\cite{Ho1996,Pu1998}. This idea was carefully analyzed in a recent theoretical work~\cite{Wolf2022} and realized shortly afterwards using a mixture of $^{87}$Rb and $^{23}$Na quantum gases \cite{Jia2022}. In this work the mechanism responsible for the emergence of the bubble shape is fundamentally different since it relies on a filled core. However it is still possible to study dynamics specific to the shell geometry, for example the expansion of a hollow bubble by removing the core just before the time-of-flight expansion, using selective radiation pressure forces~\cite{Jia2022}.

It is interesting to note that the residual inhomogeneities of the optimized dual species dipole trap used in Ref.~\cite{Jia2022} do not prevent to create a full bubble-shaped condensate for bubble trap radius of about $\SI{10}{\micro\metre}$. This technique doesn't require gravity compensation but only a matching of the trapping confinement of both species, enabled by a `magic' wavelength for the optical dipole trap \cite{Jia2022}.

\section{Possible future directions}
We conclude this perspective by pointing out possible future research directions.

\subsection{Immiscible mixtures}
The observation of shell shaped condensates in immiscible mixtures~\cite{Jia2022} opens exciting perspectives toward the study of this interesting topology. It can be anticipated that this can also be achieved with other mixtures~\cite{Ho1996,Pu1998}, even in micro-gravity~\cite{Elliott2023}, eventually giving access to new knobs, such as the control of the interactions thanks to Feshbach resonances~\cite{Chang2024}. As pointed out in the numerical study of Ref.~\cite{Wolf2022}, the phase separation leads to a change in the collective modes spectrum of the system as the interspecies scattering length is varied. The collective modes are different from those of a gas confined to a shell-shaped trap, due to the presence of the inner gas.

We envision that the coupled dynamics of the inner core and outer shell condensates will exhibit interesting phenomena. One question that arises immediately concerns the equilibrium configurations of vortices in a rotating frame: vortex lattices will form but with different parameters, due to the mass imbalance or different interaction strengths, as the vortex density scales as $M\Omega/\pi\hbar$ and the vortex core size as $\sqrt{\hbar^2/M\mu}$. Moreover, the possibility for the atoms of one species to occupy the vortex cores of the other species will likely have an impact on the vortex dynamics, possibly enabling the study of angular momentum glitches~\cite{Verma2022}.

It should be even possible to study the dynamics by forcing only one of the two species, for example by preparing one in a magnetic insensitive state and using a time-dependent magnetic field to excite the other one. This will probably require the development of robust time-dependent numerical simulation tools to solve coupled Gross-Pitaevskii equations to help interpret the experiments~\cite{Pichery2023}.

\subsection{Superfluid flows on curved surfaces}
Although the true interest of the shell-shaped condensate lies in the non trivial topology, we would like to emphasize that it is nevertheless interesting to study superfluid flows on a curved open surface. Indeed it is expected that the competition between curvature, local flows and global rotation results in new superfluid dynamics, analog to what is well known in atmospheric turbulence~\cite{Skipp2022}.
Such experiments are within reach in rf-dressed magnetic traps, by combining gravity compensation protocols~\cite{Guo2022} with the ability to set the superfluid in rotation~\cite{Guo2020,Sharma2024}. 

Among all the theoretical studies reviewed in Sec.~\ref{sec:theory}, highlighting for example vortex dynamics on curved surface, very few have studied realistic protocols to set the gas into rotation on the surface. In harmonic traps, one efficient way to nucleate vortices is to rotate a quadrupolar deformation, which couples to the quadrupole modes~\cite{Madison2000,Madison2001}. It would be interesting to study a generalization of this scenario on a curved surface (opened or closed). Similarly, phase imprinting protocols~\cite{Kumar2018} may be generalized to the shell geometry. These two methods address simultaneously both hemispheres of the bubble trap and therefore will create initial states symmetric with respect to the equator, assuming that the laser propagates along the vertical axis. It would be interesting to design original excitation protocols allowing to create arbitrary (non symmetric) vortex distributions on a shell to create highly out of equilibrium flows.

\subsection{Improving the rf-dressed shell potentials}
As discussed in Sec.~\ref{sec:challenges}, the ability to fill a full bubble in rf-dressed adiabatic potentials is currently limited by rf coupling inhomogeneities. Although it may be possible to design better antennas to reduce rf field gradients, using for example more symmetric configurations, it is also very relevant to try to mitigate these effects using an independent tool, which offers more flexibility.

One the one hand, one could be tempted to use a far detuned patterned laser beam, applying a position dependent potential on the atoms, which can be used to correct for local inhomogeneities, similarly to what has been achieved to realize homogeneous one-dimensional Bose gases in an atom-chip~\cite{Tajik2019}. Obviously, the fact that the correction has to be applied onto a closed surface makes this more challenging. However, the observed inhomogeneities~\cite{Carollo2022,Guo2022} are relatively large scale and display some symmetries, which can facilitate the implementation of these corrections.

On the other hand, it may be possible to exploit the ability to create adiabatic potentials with multiple frequencies~\cite{Bentine2017,Harte2018} to improve the homogeneity of rf dressed shell traps. Here, the idea is to use one rf field at resonance to dress the atoms and create the closed surface trap and then add one (or more) off-resonant frequencies that will locally change the potential. By playing with the polarization of the extra rf for microwave \cite{Carollo2022} fields and the antenna geometries used to generate them, it may be possible to correct some inhomogeneities, with a well defined geometry.

If necessary both methods can be combined to achieve the desired level of control.

\subsection{New proposals for the sphere topology}
Recently the authors of Ref.~\cite{Boegel2023} have shown that it was possible to adapt matterwave lensing techniques, well known in harmonic traps, to the shell geometry. The idea is to reduce the spatial expansion of the initial condensate during the time of flight expansion by applying for a short amount of time a controlled potential (or force) at the beginning of the expansion. Using a numerical simulation, they have shown that a delta-kick protocol involving the bubble potential itself could help stabilizing the shape for a hundred milliseconds. This suggest that other optimal control schemes could be designed to improve the fast manipulation of a shell-shaped condensate.

It is worth noting that almost all the works reported here concern bosonic quantum gases in shell-shaped traps. A recent theory work addressed the question of a degenerate Fermi gas trapped on the surface of a sphere \cite{He2023a}, where the interplay of rotation, curvature and BEC-BCS crossover could be investigated. Given the recent experimental successes in the study of vortex dynamics in a degenerate Fermi gas~\cite{Kwon2021}, it would be challenging but very interesting to realize the bubble trap geometry for fermions.

Finally we highlight the fact that quenches on the surface of a sphere (or on other closed surface) has been little studied. It would be certainly interesting to explore the Kibble-Zurek scenario in this geometry and compare it to the case of the ring topology~\cite{Corman2014}. With rf dressed potential it is in principle straightforward to implement an expansion quench, similar to what has been done in a ring geometry~\cite{Eckel2018}. More generally the topic of far-from-equilibrium ultra-cold atoms on closed surfaces remains largely unexplored and we can expect exciting developments in this field in the near future.

\begin{acknowledgments}
We acknowledge enlightening discussions on rf-dressed adiabatic potentials and bubble traps with Barry Garraway, Wolf von Klitzing, Chris Foot, Vanderlei Bagnato, Jook Walraven, Aiden Arnold, Anna Minguzzi, Maxim Olshanii, Mônica Caracanhas, Lucas Madeira, and Sergey Nazarenko as well as with the current and former members of the BEC group at LPL, in particular Thomas Badr, Laurent Longchambon, Vincent Lorent, Aur\'elien Perrin and Paul-\'Eric Pottie. We also thank Fabrice Wiotte for the development of low noise DDS rf sources. LPL is UMR 7538 of CNRS and Universit\'e Sorbonne Paris Nord.
This work has been supported by R\'egion \^Ile-de-France in the framework of DIM SIRTEQ (project Hydrolive).
We acknowledge financial support from the ANR project VORTECS (Grant No. ANR-22-CE30-0011).
\end{acknowledgments}


%
\end{document}